\documentclass[prl,aps,twocolumn,superscriptaddress,showpacs,floatfix,letterpaper]{revtex4-1}  
\usepackage{hyperref}
\usepackage{amssymb}
\usepackage{amsmath}
\usepackage{float}
\usepackage{bbm}
\usepackage{graphicx}
\usepackage{epsfig}
\usepackage{epstopdf}
\usepackage[usenames,dvipsnames]{color}
\usepackage{soul}
\usepackage{bibunits}

\newcommand{\omo}{\Omega_0}

\hypersetup{
  colorlinks = true, 
  urlcolor   = blue, 
  linkcolor  = blue, 
  citecolor  = blue  
}
\begin{document}
 
\title{Band Structure and Terahertz Optical Conductivity of Transition Metal Oxides: Theory and Application to CaRuO$_3$}

\author{Hung T. Dang}
\affiliation{Institute for Theoretical Solid State Physics, JARA-FIT and JARA-HPC, RWTH Aachen University, 52056 Aachen, Germany}
\author{Jernej Mravlje}
\affiliation{Jo\v{z}ef Stefan Institute, Jamova 39, SI-1000, Ljubljana, Slovenia}
\author{Antoine Georges}
\affiliation{Coll{\`e}ge de France, 11 place Marcelin Berthelot, 75005 Paris, France}
\affiliation{Centre de Physique Th\'eorique, Ecole Polytechnique, CNRS, 91128 Palaiseau Cedex, France}
\affiliation{DQMP, Universit{\'e} de Gen{\`e}ve, 24 quai Ernest-Ansermet, 1211 Gen{\`e}ve 4, Switzerland}
\author{Andrew J. Millis}
\affiliation{Department of Physics, Columbia University, New York, New York 10027, USA}
\date{\today}

\begin{abstract}
Density functional plus dynamical mean field calculations are used to show that in transition metal oxides, rotational and tilting (GdFeO$_3$-type) distortions of the ideal cubic perovskite structure  produce a multiplicity of  low-energy optical transitions which affect the conductivity down to  frequencies of the order of $1$ or $2$~mV (terahertz regime), mimicking non-Fermi-liquid effects even in systems with a strictly Fermi-liquid self-energy. For CaRuO$_3$, a material whose measured electromagnetic  response in the terahertz frequency regime has been interpreted as evidence for non-Fermi-liquid physics, the combination of these band structure effects and  a renormalized Fermi-liquid self-energy  accounts for the  low frequency optical response which had previously been regarded as a signature of exotic physics. Signatures of deviations from Fermi-liquid behavior at higher frequencies ($\sim 100$~meV) are discussed.
\end{abstract}

\maketitle

\begin{bibunit}

Fermi-liquid theory provides the canonical picture of metals; the observation of deviations from Fermi-liquid behavior is thus of intense interest as a potential indication of novel physics. The defining feature of a Fermi-liquid is the existence of electronlike quasiparticles whose low temperature and frequency properties are characterized by an effective mass that is independent of frequency  and a scattering rate that is parametrically smaller than frequency or temperature (typically varying as $\omega^2$ or $T^2$).  The perovskite ruthenate CaRuO$_3$ has been the subject of considerable attention in this context because its frequency-dependent conductivity has been reported \cite{Lee02,Kamal06} to vary as a  power of frequency with an exponent less than unity. The anomalous dependence extends to very low frequencies of the order of $1$~THz ($\sim 4$~meV) \cite{Kamal06}, and this has been interpreted as indicating a breakdown of Fermi-liquid physics in this material. Similar interpretations have been given of optical data in SrRuO$_3$ \cite{Kostic98}. On the other hand, recent dc transport measurements in CaRuO$_3$ found quantum oscillations and a quadratic temperature dependence of the resistivity \cite{Schneider14} below $1.5$~K -- characteristic of a Fermi-liquid. The link between the frequency-dependent and dc transport measurements has not been established and a model accounting for the optical conductivity is not known. 

In this Letter we present density functional theory (DFT) plus dynamical mean field theory (DMFT) calculations which indicate that band structure effects associated with octahedral rotations of the ideal perovskite crystal structure  produce  optically active interband transitions that contribute to the conductivity on scales as low as $1$~THz ($\sim$~meV) and can mimic non-Fermi-liquid physics.   As an application we show that the observed terahertz conductivity of CaRuO$_3$ is consistent with Fermi-liquid-like quasiparticles and quantify the departures from Fermi-liquid physics that occur at higher scales.

The standard arguments relating optical conductivity $\sigma(\Omega)$  and electron self-energy $\Sigma(\omega)$ may conveniently be framed in terms of an approximation  due to Allen \cite{Allen71,Allen04,Basov11}:
\begin{equation}
\sigma(\Omega)\propto\frac{i}{\Omega}\int d\omega\frac{f(\omega)-f(\omega+\Omega)}{\Omega-\Sigma(\Omega+\omega)+\Sigma^{\star}(\omega)}.
\label{sigmaAllen}
\end{equation}
Here $f$ is the Fermi function and the $^\star$ denotes complex conjugation. Equation~\eqref{sigmaAllen} is expected to be  reasonable  when interband transitions are neglected.

In a simple Drude metal, $\Sigma=-i/(2\tau)$, with $2\tau$ representing the time between scatterings of electrons off of impurities. Use of this self-energy in  Eq.~\eqref{sigmaAllen} yields the familiar Drude conductivity  $ \sigma(\Omega)\propto \tau/(1-i\Omega\tau)$. Use of the Fermi-liquid form $\Sigma(\omega) \propto (1-Z^{-1})\omega - i \Omega_0^{-1}\left( \omega^2 +\pi^2 T^2\right)$   yields a conductivity with a characteristic scaling form~\cite{Berthod13} (see also Refs.~\cite{Stricker14,Chubukov-2012}) that we will refer to as the single-band Fermi-liquid (SBFL) conductivity. If the self-energy takes the non-Fermi-liquid form $\Sigma(\omega)\sim \omega^x$ with $x<1$, one has $|\Sigma(\omega)|>|\omega|$ at low frequency, so that  the term proportional to $\Omega$ in the denominator of  the argument of the integral in Eq.~\eqref{sigmaAllen} may be neglected.  A scaling analysis of Eq.~\eqref{sigmaAllen} then shows that  for small $\Omega$, $\sigma\sim \Omega^{-x}$, with the divergence cut off by temperature.  

We compare expectations based on Eq.~\eqref{sigmaAllen} to realistic calculations of the  frequency-dependent conductivity of CaRuO$_3$. This material crystallizes in a GdFeO$_3$-distorted version of the ideal cubic perovskite structure. In the latter, there are three  near-Fermi-surface bands derived from the three $t_{2g}$ states.  The GdFeO$_3$-distorted structure has four Ru ions in the unit cell, leading to 12 $t_{2g}$-derived near-Fermi-surface bands. The $t_{2g}$-derived bands are the eigenvalues of  a Hamiltonian matrix $H_0(\mathbf{k})$ with $\mathbf{k}$ being a wave vector in the first Brillouin zone. $H_0(\mathbf{k})$ is obtained by using maximally localized Wannier function (MLWF) \cite{Marzari97,Souza01} techniques  to project the Kohn-Sham Hamiltonian found from a spin-unpolarized DFT band calculation onto the near-Fermi-surface states. The effects of electron-electron interactions are encoded in the self-energy $\Sigma(\mathbf{k},\omega)$, also a matrix, which we compute by applying  single-site DMFT to the $H_0(\mathbf{k})$ corresponding to the experimental structure of CaRuO$_3$  with standard Slater-Kanamori interactions parametrized by $U=2.3$~eV and $J=0.4$~eV  (see the Supplementary Material \cite{supp} for details of DFT and DMFT calculations). Electron propagation in the $t_{2g}$-derived bands is  thus described by the $N\times N$ matrix Green function ($N=3$ for the cubic structure and $12$ for the experimental one) 
\begin{equation}
 \mathbf{G}(k,\omega)=\left[\omega+\mu-H_0(\mathbf{k})-\Sigma(\omega)\right]^{-1}.
 \label{G}
 \end{equation}
 
For the situations we consider there are no vertex corrections to the current operator in the single-site dynamical mean field approximation, essentially because no on-site optical transitions are allowed (see the Supplementary Material \cite{supp} for a more detailed discussion and also Ref.~\onlinecite{Khurana90} for the single-band case), so the conductivity becomes
\begin{eqnarray}
\sigma(\Omega)&=&\int\frac{d\omega}{\pi}\frac{f(\omega)-f(\omega+\Omega)}{\Omega}
\label{sigmadef} 
\\
&&\times \mathrm{Tr}\left[J_\mathbf{k} \mathrm{Im}G(\mathbf{k},\omega+\Omega) J_\mathbf{k} \mathrm{Im}G(\mathbf{k},\omega)\right].
\nonumber
\end{eqnarray}
The matrix current operator $J_\mathbf{k}$ is  derived in a standard way from $H_0(\mathbf{k})$ (note that in systems with more than one atom per unit cell, care must be taken to use a basis in which each atom acquires the Peierls phase appropriate to its physical position within the unit cell \cite{Wang11,Tomczak09}). The trace is over momentum and band indices. In our calculations, the four-dimensional integral (in frequency and momentum space) is performed using a Gaussian quadrature with $60$ points in each direction. The Allen formula [Eq.~\eqref{sigmaAllen}] may be derived from Eq.~\eqref{sigmadef} if the matrices are diagonal (no interband transitions) and when the transport function (i.e. the density of states weighted by current matrix elements) depends weakly on energy. 

\begin{figure}[t]
  \includegraphics[width=\columnwidth]{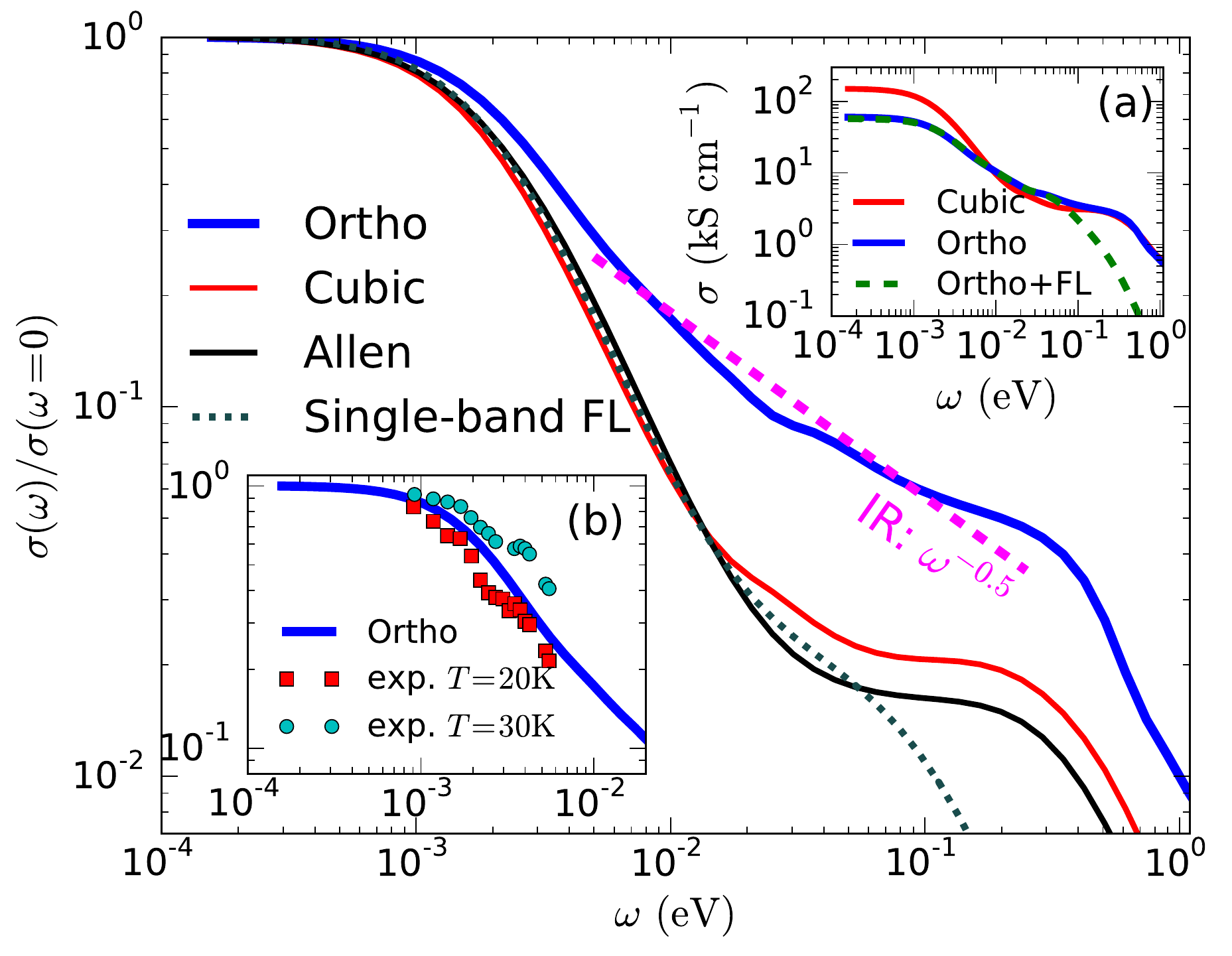}\\
\caption{\label{fig:conductivities1}(Color online) 
Main panel: Optical conductivity (normalized to its zero frequency value). (Heavy solid blue line) Conductivity computed from DFT+DMFT at $T=30$~K for the orthorhombic experimental structure. The dashed straight line is a guide to the eye  indicating the power-law behavior $\sim \omega^{-0.5}$, corresponding to the experimentally reported mid-IR frequency dependence \cite{Lee02}. Also displayed are the conductivities computed for hypothetical cubic structure (intermediate weight red line)  and from the Allen formula (light black line), using the same DMFT self-energy as was used  in the experimental structure calculation. Finally, the dotted  line presents the `SBFL' result obtained by using a Fermi-liquid self-energy in the Allen formula.
[Inset (a)] Optical conductivity calculated using DMFT self-energy for experimental  structure  and cubic structure compared to a calculation (dashed lines) for the experimental structure but with a  Fermi-liquid self-energy [Eq.~\eqref{sigmafl}]. [Inset (b)] Experimental data of Ref.~\onlinecite{Schneider14} in the terahertz range along with the DMFT calculation for the realistic structure.
}
\end{figure}

The main panel of Fig.~\ref{fig:conductivities1} presents the normalized conductivity calculated using Eq.~\eqref{sigmadef} with $H_0(\mathbf{k})$, $J_\mathbf{k}$ and $\Sigma$ appropriate to the experimental structure of CaRuO$_3$. The conductivity in the midinfrared regime ($5~\mathrm{meV}\lesssim \Omega\lesssim 250~\mathrm{meV}$) appears to vary as a power law $\sim \omega^{-x}$ with $x$ being in the range $0.4-0.6$, similar to the power law reported experimentally~\cite{Lee02}. The lower inset compares the calculated conductivity to recent measurements~\cite{Schneider14}, which come from samples with significantly lower impurity scattering than samples studied earlier~\cite{Kamal06}. The quantitative correspondence between calculation and data is good.

Also shown in Fig.~\ref{fig:conductivities1} are the conductivities obtained from Eq.~\eqref{sigmadef} using the $H_0(\mathbf{k})$ and $J_\mathbf{k}$ corresponding to the ideal cubic structure (while keeping the same self-energies as for the real structure) and by using the Allen formula [Eq.~\eqref{sigmaAllen}] again for the same self-energies.  (The Allen formula results are obtained as an equally weighted sum over three terms, one for each diagonal entry in the self-energy matrix).  In the terahertz and subterahertz regime ($\omega \sim 1-10$~meV) the cubic or Allen results exhibit a much more rapid roll-off from the dc plateau than does the experimental structure conductivity, while in the mid-IR ($\omega \sim 100$~meV) regime the cubic or Allen results exhibit an approximate plateau if the DMFT self-energy is used.

\begin{figure}[t]
\includegraphics[width=\columnwidth]{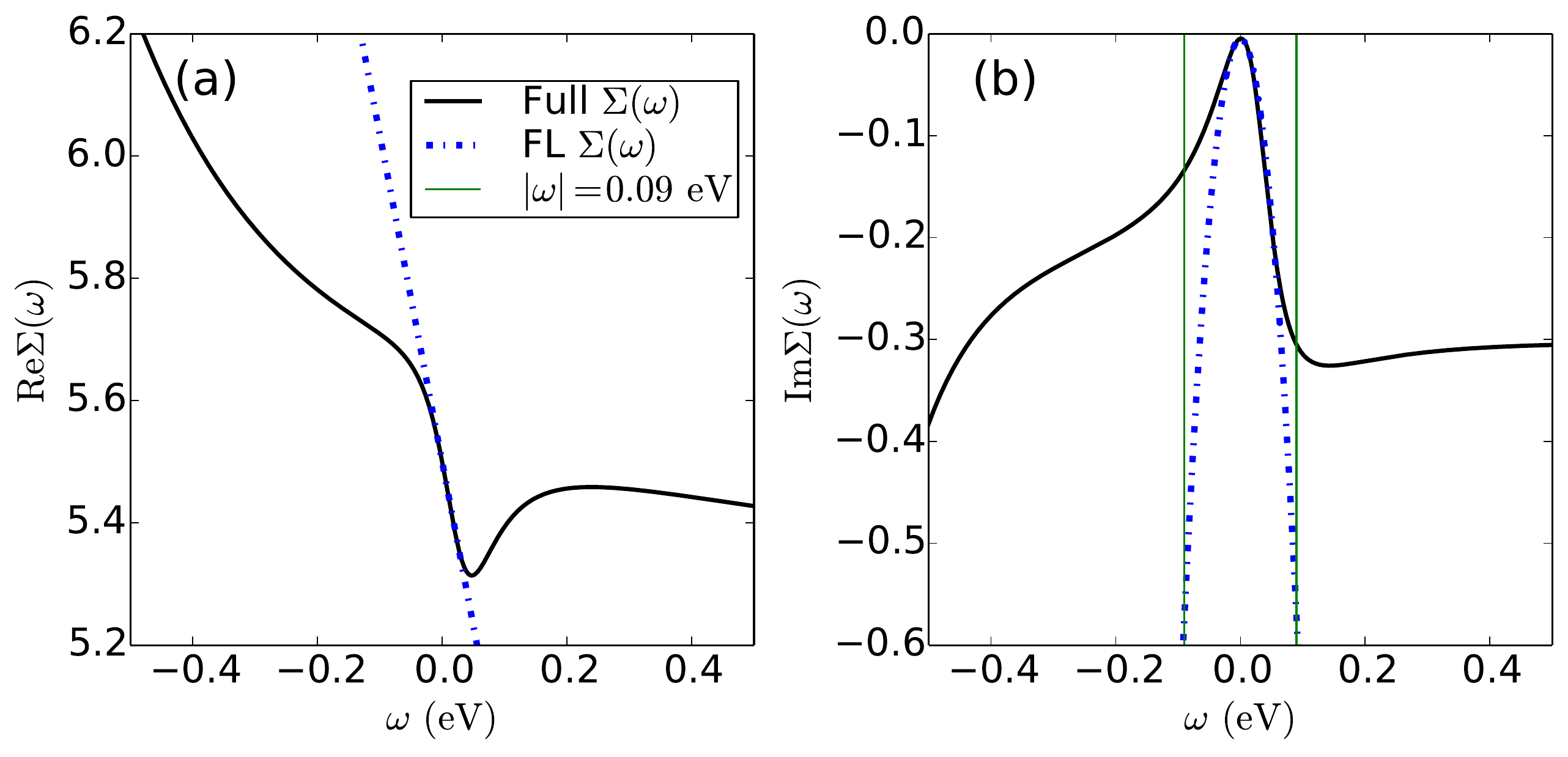}
\caption{\label{fig:selfenergies}(Color online) (a) Real part of the self-energy for one of the three orbitals (solid line). (Dash-dotted line) Linear low-frequency fit to the real part of Eq.~\eqref{sigmafl} with slope $1-Z^{-1}\equiv d\Sigma/d\omega=-5.3$. (b) Imaginary part of self-energy of the same orbital (solid line). (Dash-dotted line) Low-energy Fermi-liquid fit to the imaginary part of Eq.~\eqref{sigmafl} with $\omo\simeq 14$~meV and temperature $T=0.0025~\mathrm{meV}\approx 30$~K. (Vertical lines) Boundary of the Fermi-liquid region ($\omega=0.09$~eV).}
\end{figure}

As the self-energies used in the cubic and orthorhombic calculations are exactly the same, the difference between the results is not a self-energy effect and does not correspond to non-Fermi-liquid physics. To probe the effect of the self-energy, we also display in the main panel the SBFL result (with parameters $Z, \Omega_0$ determined from a fit to the DMFT self-energy in Fig.~\ref{fig:selfenergies}), and in the inset of Fig.~\ref{fig:selfenergies}(a) the conductivity obtained when using a Fermi-liquid self-energy and the realistic orthorhombic structure. We see that the choice of self-energy hardly influences the low frequency result; it is only at frequencies higher than $\sim 50$~meV that the choice of self-energy significantly influences the calculation. 

\begin{figure}[t]
 \includegraphics[width=0.95\columnwidth]{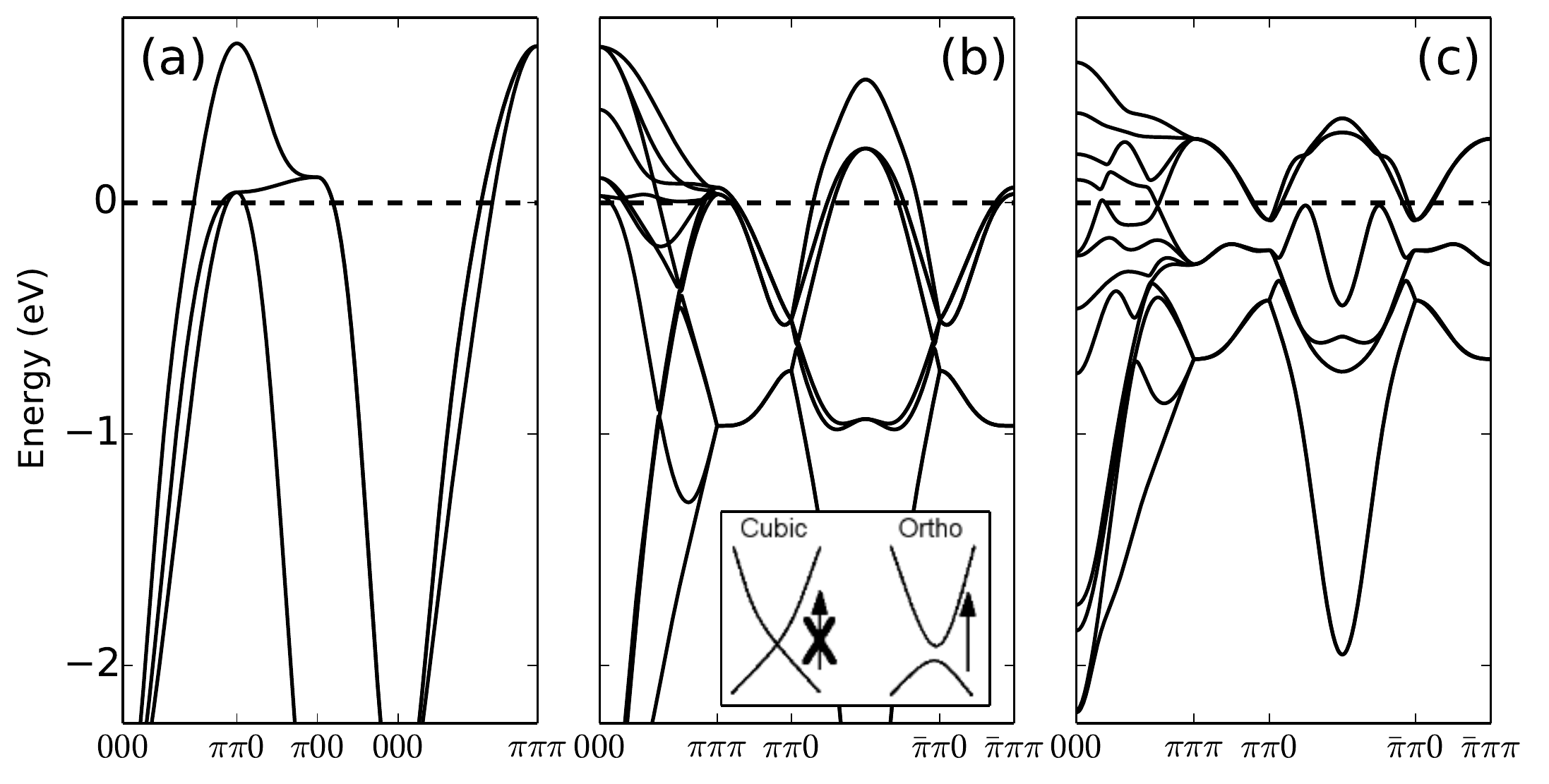}
\caption{\label{fig:band_structure} Band structure computed for
(a)  the ideal cubic perovskite form of CaRuO$_3$ plotted along high symmetry directions in the cubic perovskite Brillouin zone;
(b)  the ideal cubic structure folded back into the Brillouin zone of the experimental orthorhombic structure;
and (c) the experimental orthorhombic structure, along high symmetry  directions of the orthorhombic Brillouin zone. 
All three panels show the frontier $t_{2g}$-antibonding bands produced by a MLWF fitting of the DFT (generalized gradient approximation) band structure.
(Inset) Optical transitions across minigaps which are forbidden in the cubic structure are activated in the distorted structure. 
}
\end{figure}

The key approximation of both the Allen and SBFL formulas is the neglect of interband transitions. The difference between these approximations and the calculation for the experimental structure thus arises from optically active interband transitions, which  are seen to affect the conductivity down to scales as low as $1$~THz. To explicate the origin of these transitions, we present in Fig.~\ref{fig:band_structure} our calculated band structures. The left panel shows the  near-Fermi-surface bands found for the  ideal cubic structure.  Direct interband transitions between the three bands are possible in principle; however, the different orbital symmetry of the different bands means that the matrix elements are  small, especially in the lower frequency regime, so that the cubic and Allen results are similar. At higher energies, interband transitions have some effect in the cubic structure as well.  

The middle panel shows the bands of the cubic structure, folded into the Brillouin zone of the experimental structure. The backfolding creates the possibility of many low-lying interband transitions, but in the cubic structure these transitions are not optically active, as they do not correspond to zero-momentum transfer. The right panel shows the band structure obtained for the  experimental structure. The octahedral rotations reduce the overlap between states on different sites, causing a  band narrowing from $3.6$ to $2.6$~eV visible, for example, in the energies near the $\Gamma$ point and flattening the dispersion in the near-Fermi-surface region.  The zero frequency conductivity of the orthorhombic case is thus  smaller (by a factor of $\sim 3$) than the cubic result (see the inset in Fig.~\ref{fig:conductivities1}). The rotations also allow matrix elements between nearby states, opening additional minigaps where the cubic bands cross, further flattening the bands at the Fermi level and, crucially, activating optical matrix elements between the backfolded bands.  Starting at $\omega \sim 1$~meV, these become important, changing the functional form of the conductivity. At high frequencies the experimental and cubic structure conductivities become very similar, as the small gaps are unimportant. 

To study the nature of the non-Fermi-liquid effects in the conductivity of CaRuO$_3$ we present in  Fig.~\ref{fig:selfenergies} a plot of the real and imaginary parts of the self-energy calculated for one of the three $t_{2g}$ orbitals (the self-energies associated with the other two are similar).  Also shown is a fit of the self-energy to the functional form
\begin{equation}\label{sigmafl}
 \Sigma_{FL}(\omega,T)=\left(1-Z^{-1}\right)\omega-i\,\omo^{-1}\left[\omega^2+b\left(\pi T\right)^2\right].
\end{equation}
Here  $Z$ is a dimensionless constant giving the mass renormalization $m^\star/m\equiv Z^{-1}$, $T$ is the temperature and the characteristic energy $\omo$ sets the scale of the scattering rate. The parameter $b=1$  (Fermi-liquid) for the plotted orbital but about $2$ to $3$ for the other two perhaps because the coherence temperature is not quite reached.  From  Fig.~\ref{fig:selfenergies} we see that at very low frequencies  $|\omega|\lesssim 20$~meV the self-energy approximately takes the Fermi-liquid form, but for larger frequencies $\omega \gtrsim 40$~meV pronounced ($\gtrsim 50\%$) deviations occur. On the positive frequency (electron addition) side the imaginary part of the self-energy saturates for $\omega \gtrsim 0.1$~eV and the real part loses most of its frequency dependence. On the negative frequency (electron removal) side the imaginary part of the self-energy increases (although not as rapidly as the Fermi-liquid  $\omega^2$) and exhibits a large peak (not shown)  at $\omega \sim -1$~eV. The low frequency at which deviations from Fermi-liquid behavior occur is characteristic of multiorbital systems with a sizeable Hund's coupling \cite{Georges13}. 

At very low frequencies ($\lesssim 7.5$~meV), the frequency $\omega$ is less than $\pi T$ so the scattering rate is in effect constant.  The cubic and Allen formula conductivities in Fig.~\ref{fig:conductivities1} are indeed well described by a Drude form with frequency independent scattering rate $\Gamma_\mathrm{Drude}=2Z(\pi T)^2/\omo \approx 1.4$~meV.  However, in the orthorhombic structure  interband transitions cause the conductivity  to decay much less rapidly than expected from the Drude formula at frequencies  $\gtrsim 1$~meV.

Suppose now that the self-energy was well described by the Fermi-liquid form even at frequencies higher than $\sim 20$~meV. Inspection of the upper inset of Fig.~\ref{fig:conductivities1} shows that for $\Omega \gtrsim 100$~meV the corresponding conductivity becomes much smaller than either the cubic or the experimental system conductivity. [One sees this from the behavior of the SBFL curve in the main panel of Fig.~\ref{fig:conductivities1}, but it can also be derived directly (see the Supplementary Material \cite{supp})  by inserting  Eq.~\eqref{sigmafl} into Eq.~\eqref{sigmaAllen},  setting temperature $T=0$ and scaling the internal integration variable by the external frequency $\Omega$ to yield
$
\sigma_\mathrm{FL}(\Omega)\propto 
Z\,\int_{-1}^0 dx
\left(-i\Omega+Z\omo^{-1}\Omega^2 (1+2x+2x^2)\right)^{-1}.
$
The real part of this expression has an approximately Lorentzian Drude-like decay with decay constant $\Omega_0/Z\approx 80$~meV, which describes well the high-frequency behavior of the Fermi-liquid results.) The slower decay of the actual conductivity is a signature of  deviations from Fermi-liquid physics. It results in particular from the saturation of the scattering rate and the strong deviation of   $\mathrm{Re}\Sigma(\omega)$ \cite{Mravlje11} from the low frequency linear behavior apparent in Fig.~\ref{fig:selfenergies}. In this non-Fermi-liquid higher-frequency regime, the similarity of the conductivities for the cubic and experimental structure conductivities shows that band structure effects are of less importance here, implying that information about the self-energy may be extracted from the conductivity.

\begin{figure}[t]
\includegraphics[width=\columnwidth]{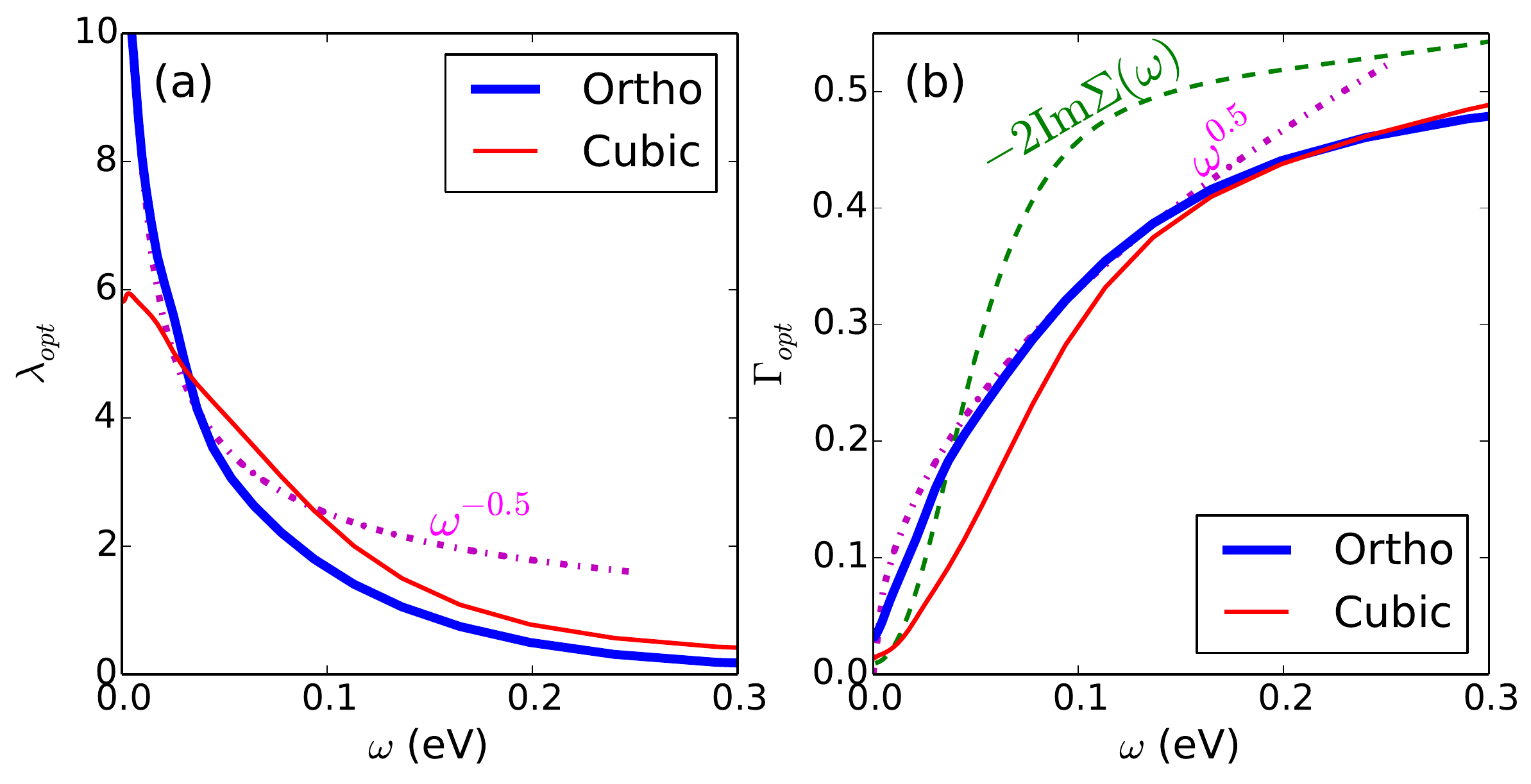}
\caption{\label{fig:memory1}(Color online) Optical mass (a) and scattering rate (b) obtained via Eq.~\eqref{Mdef} for different cases considered in this Letter and compared to quasiparticle mass [$\sim 6.7$ as calculated from the slope of $\mathrm{Im}\Sigma(i\omega_n)$] and the imaginary part of the single particle self-energy (dashed green curve). Dotted magenta curves indicate $\omega^{\pm0.5}$ behavior.}
\end{figure}

Formally inverting Eqs.~\eqref{sigmaAllen} or \eqref{sigmadef} to obtain self-energies from measured conductivities is an ill-posed and essentially unsolvable inversion problem. However the widely used ``memory function'' method \cite{Basov05} provides considerable insight.   The typical procedure is to express the complex conductivity $\tilde{\sigma}$  in terms of an optical mass enhancement $\lambda_\mathrm{opt}$ and scattering rate $\Gamma_\mathrm{opt}$ defined   as
\begin{equation}
\tilde{\sigma}=\frac{K}{ -i\omega\left(1+ \lambda_\mathrm{opt}(\omega)\right)+\Gamma_\mathrm{opt}(\omega)}.
 \label{Mdef}
 \end{equation}
The objects $\lambda_\mathrm{opt}$ and $\Gamma_\mathrm{opt}$  are often interpreted as mass enhancement and scattering rate, respectively, and are assumed to provide information about the electron self-energy. Their frequency dependence is determined by the frequency dependence of the complex conductivity while the overall magnitude is determined by the constant $K= 2/\pi\int_0^\infty \mathrm{Re}\tilde{\sigma}(\omega) d\omega$.  In the two panels of  Fig.~\ref{fig:memory1} we present the $\lambda_\mathrm{opt}$ and $\Gamma_\mathrm{opt}$ determined from our calculations, using the directly computed sum rule values $K_\mathrm{cubic}=0.165$~eV and $K_\mathrm{ortho}=0.153$~eV (computed for the orthorhombic $b$ direction).  

For the cubic and Allen-formula cases, where interband transitions are not important, the scattering rate found from the memory function is in reasonable  agreement with a particle-hole average of an imaginary part of the self-energy. The  scattering rate magnitude is correctly estimated and the low frequency $\omega^2$ behavior is clear. The low frequency limit of the mass corresponds precisely to the quasiparticle mass enhancement and the decrease of mass at higher frequency reflects the flattening of the $\mathrm{Re}\Sigma$ curve \cite{Stricker14}. 

For calculations performed with the experimental structure the  situation is different: at low frequencies the inferred scattering rate is too large by a factor  $\gtrsim 2-4$ and  has the wrong concavity. In fact, the inferred scattering rate is roughly consistent with an $\omega^{1/2}$ behavior and, similarly, over a limited low frequency range the optical mass can be fit as $\omega^{-1/2}$. This  suggests that caution is warranted in performing a memory function analysis of the low frequency data on GdFeO$_3$-distorted materials. However, the reasonable correspondence at higher frequencies ($\omega \gtrsim 100$~meV) between the optical scattering rate and the imaginary part of the self-energy (averaged over positive and negative frequencies) again confirms that in this range the conductivity does give a reasonable estimate of the magnitude and the saturation frequency of the self-energy, and in this sense reveals non-Fermi-liquid behavior of the Hund's metal kind. 

In summary, using CaRuO$_3$ as an example  we have shown that real materials effects, in particular a multiplicity of optically allowed low-lying transitions arising from band folding due to rotational and tilt distortions, can produce a low frequency conductivity of the form previously associated with non-Fermi-liquid physics. A direct diagnosis of universal Fermi-liquid behavior from the optical conductivity, along the lines of Ref.~\onlinecite{Berthod13}, only applies when such effects are not important. Our results call for a reexamination of other reports of unusual optical response, for instance in  SrRuO$_3$, which has a ferromagnetic ground state and a smaller orthorhombic distortion, and for which ARPES spectra consistent with Fermi-liquid behavior are observed \cite{Shai13}. It is also important to examine whether low-lying interband transitions complicate the analysis \cite{Sulewski88,Katsufuji99,Yang06,Nagel-2012,Mirzaei13}  of the ratio of the $T^2$ and $\omega^2$ terms in the optical scattering rate, which has been argued to be inconsistent with Fermi-liquid theory.  

\begin{acknowledgments}
H.T.D. acknowledges support from the Deutsche Forschungsgemeinschaft (DFG) within Projects No. FOR 1807 and No. RTG 1995, as well as the allocation of computing time at J\"ulich Supercomputing Centre and RWTH Aachen University through JARA-HPC. J.M. acknowledges support of the Slovenian research agency under the program P1-0044. A.G. acknowledges a grant from the European Research Council (ERC-319286 QMAC) and the Swiss National Science Foundation (NCCR-MARVEL). The work of A.J.M. on this project was supported by the US National Science Foundation under grant DMR-1308236.
\end{acknowledgments}

\end{bibunit}

%
%
%

\begin{bibunit}
\clearpage 
\setcounter{figure}{0}
\renewcommand{\thefigure}{\arabic{figure}S}
  
\makeatletter
     \@addtoreset{figure}{section} 
\makeatother

\onecolumngrid
\begin{center}
\Large{\bf Supplementary Material for \\ Band Structure and Terahertz Optical Conductivity of Transition Metal Oxides: Theory and Application to CaRuO$_3$}
\end{center}
\twocolumngrid

 \date{\today}

\section{Density Functional Calculations}

\begin{figure}[b]
  \includegraphics[height=0.3\textheight]{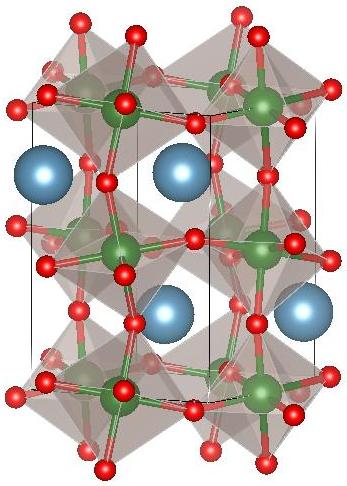}
\caption{\label{fig:cro_crystal_structure} (Color online) The crystal structure of CaRuO$_3$ using atomic position data from Ref.~\onlinecite{Bensch90}. Largest balls (blue online) are Ca, balls of intermediate size inside the octahedra (green online) are Ru, and smallest balls at the vertices of the octahedra (red online) are oxygen. The plot is generated by {\sc VESTA} program \cite{Momma11}.}

\end{figure}

We obtained the Hamiltonian matrix $H_0(\mathbf{k})$ and the current operator $J_\mathbf{k}$  for both the  ideal cubic perovskite structure (lattice constant $a_0=3.84$~\AA) and  the experimentally observed structure (atomic positions from Ref.~\onlinecite{Bensch90} - see Fig.~\ref{fig:cro_crystal_structure} for the lattice structure) by using maximally localized Wannier function techniques \cite{Marzari97,Souza01} to project the Kohn-Sham Hamiltonian found from a spin-unpolarized density functional theory (DFT) band calculation onto the near-Fermi-surface states. The DFT calculations were performed using {\sc Quantum ESPRESSO} \cite{QE-2009,QEPseudo} with a $k$ mesh of  $13\times13\times13$, cutoff frequency $30$~Ry and the Perdew-Burke-Ernzerhof exchange correlation functional \cite{Perdew96}.

The projection to the near-Fermi-surface bands was performed using the  {\sc wannier90} code \cite{Mostofi08}. For the experimental structure, the energy window for the projection is from $-3$ to $1$~eV; for the ideal cubic structure, it ranges from $-4$ to $2.6$~eV. We find that in the ideal cubic structure the Ru $d$ $t_{2g}$-derived  bands overlap  the oxygen bands in the region of the zone center (energies $2$~eV or more below the Fermi level), leading to  imperfections in the Wannier fit in these regions, especially in the one Ru per unit cell calculation. These small details relate to states several eV away from the Fermi surface and do not affect our conclusions, which concern excitations with energies $\lesssim 1$~eV.

\section{Dynamical Mean Field Theory Calculations}
 
We obtained the self-energy $\Sigma(\omega)$ from  single-site dynamical mean field (DMFT) calculations where the correlated orbitals are the $t_{2g}$-derived bands of the $H_0(\mathbf{k})$ corresponding to the experimental structure of CaRuO$_3$ and the additional interactions are of the rotationally invariant Slater-Kanamori form \cite{Imada98} with $U=2.3$~eV and $J=0.4$~eV. The $U$ and $J$ parameters are  close to those found in previous studies for Sr$_2$RuO$_4$ \cite{Mravlje11,Vaugier12} and very recent studies of   SrRuO$_3$ and CaRuO$_3$ \cite{Dang_srocro}. However, the precise values are not important here: our goal is not to model the materials in quantitative detail but rather to uncover general features.

Our DMFT calculations use the hybridization expansion version of continuous-time quantum Monte Carlo method \cite{Werner06} implemented in the TRIQS package \cite{triqs_project},  and temperature set to $2.5~\mathrm{meV}\sim30$~K. The momentum integral in the DMFT self-consistent loop is calculated using Gaussian quadrature with $26$ points in each direction. The resulting imaginary-time self-energies are continued using the Pade method \cite{Vidberg77}.  

In the DMFT approximation the self-energy is local so $\Sigma$ is block diagonal in the Wannier basis of orbitals localized on the Ru sites. On each site an orbital basis may be found for which the self-energy is, to a very good approximation, orbitally diagonal.  In all of the cases we have considered this basis is the one found by the {\sc wannier90} code, and is  such that for each Ru site, the local $\hat{x}, \hat{y}$ and $\hat{z}$ axes point towards the three different directions of the RuO$_6$ octahedron. Within this basis, the off-diagonal elements of the Hamiltonian matrix contribution is small, less than 10\% of the matrix norm. (We have tested that using a more optimal choice of basis with off-diagonal contribution of about 5\% of the matrix norm does not change the results appreciably.) A simple rotation relates the orbital basis on one Ru site in the unit cell to that in another. The important part of the self-energy is thus a $3\times 3$ diagonal matrix and we neglect the off-diagonal terms in our calculations. The three diagonal components of the self-energies turn out to be similar.  

\section{Optical conductivity calculations}

The optical conductivity $\sigma(\Omega)$ [Eq.~(3) in the manuscript] is a four-dimensional integral in frequency and momentum space. We calculate $\sigma(\Omega)$ by using Gaussian quadrature with 60 points in each direction. For the frequency integration, because of $\frac{f(\omega)-f(\omega+\Omega)}{\Omega}$ in $\sigma(\Omega)$, we set the integrating interval for $\omega$ from $-W_F T - \Omega$ to $W_F T$ with the Fermi width $W_F=10$ (reminding that $T=2.5~\mathrm{meV}\approx 30$~K in our calculations and $f(\omega)$ is the Fermi-Dirac distribution). 

\begin{figure}
\centering
 \includegraphics[width=\columnwidth]{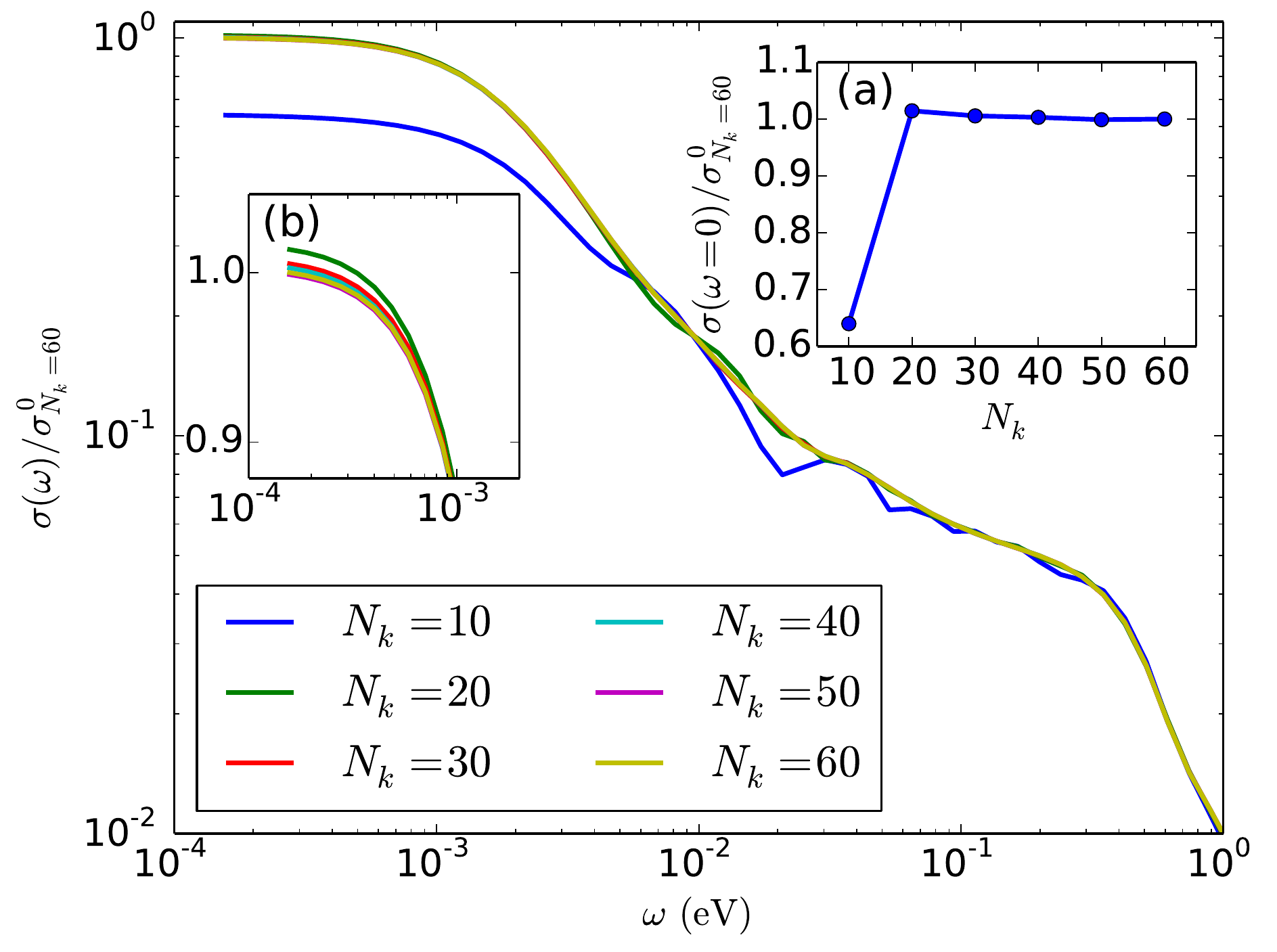}
\caption{\label{fig:optics_kpoints} Calculations for the optical conductivity of distorted CaRuO$_3$ at $T=0.0025$~eV, $U=2.3$~eV and $J=0.4$~eV. Main panel: optical conductivity in the log-log scale plotted using different number of points per direction $N_k$ for the 4$d$ integrals in a wide energy range. Inset (a): the DC conductivity ($\sigma(\omega=0)$) vs. number of integral points per direction. Inset (b): expanded plot for the optical conductivity in main panel for a narrow energy range. All the plots are rescaled by the DC conductivity value obtained from the integral using 60 points per direction.}
\end{figure} 

Using $60$ points per direction is already good enough to justify the convergence of the calculations. We have checked for some cases by comparing optical spectra produced using different numbers of points per direction. Figure~\ref{fig:optics_kpoints} demonstrates the dependence of the result on the number of integral points for the case of distorted structure CaRuO$_3$. It shows that the result converges quickly as the number of points per direction $N_k$ reaches 60. The optical spectra are reasonably converged by  $N_k=20$ and the DC conductivity converges at $N_k>30$.

\section{The self-energy}

\begin{figure}
\centering
 \includegraphics[width=\columnwidth]{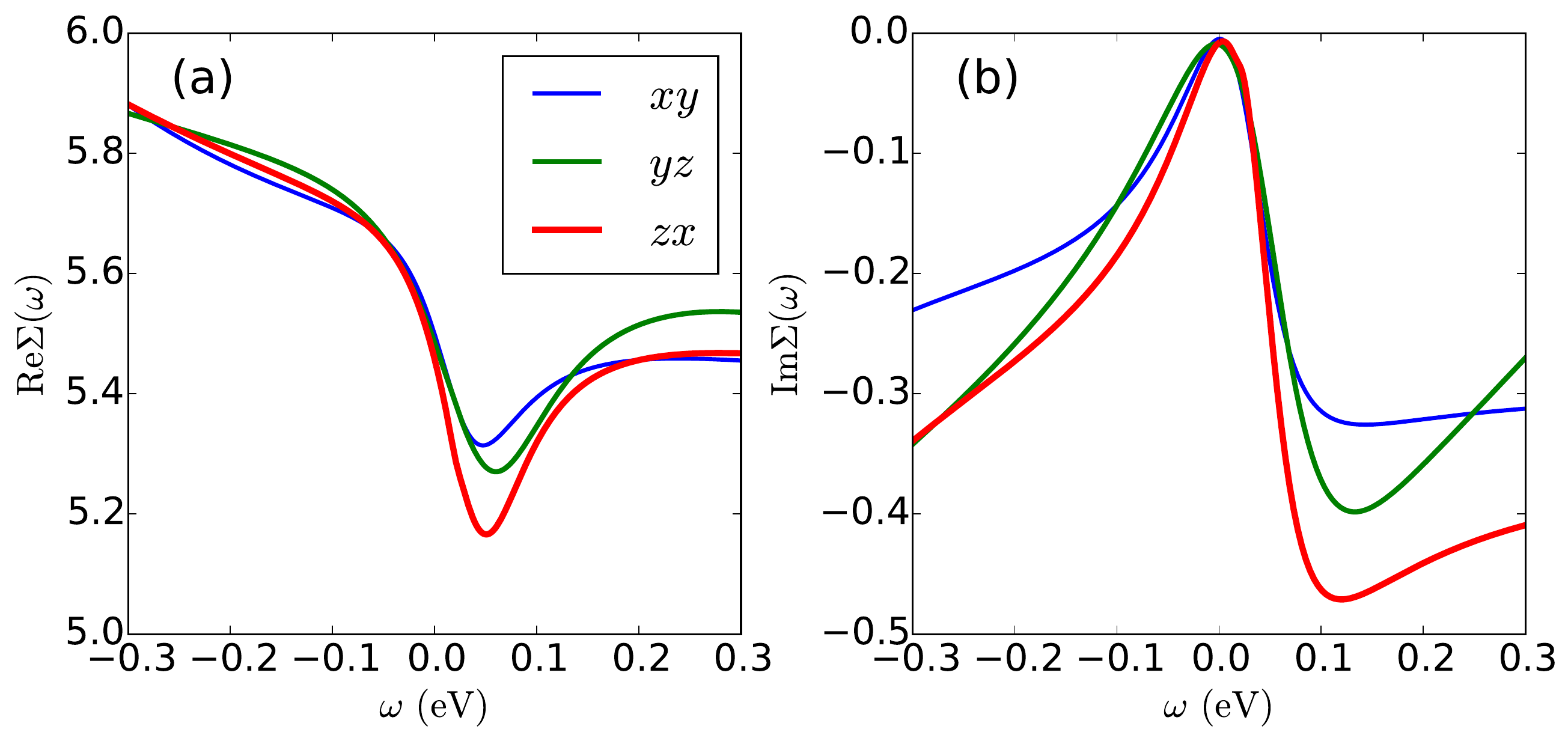}
\caption{\label{fig:self_energy_all} The  orbitally-resolved self-energy of CaRuO$_3$ calculated with $U=2.3$~eV, $J=0.4$~eV and $T=2.5$~meV. The $xy$ self-energy is  shown in the Fig.~3 of the main text.}
\end{figure}

Here for completeness we show the orbital dependence of the self-energy in Fig.~\ref{fig:self_energy_all}. 

Fitting the low frequency behavior of the self-energy to the form
\begin{equation}\label{sigmafl}
 \Sigma_{FL}(\omega,T)=\left(1-Z^{-1}\right)\omega-i\,\omo^{-1}\left[\omega^2+b\left(\pi T\right)^2\right],
\end{equation}
yields $\Omega_0 = 0.014,~0.022,~0.016$~eV, and $b=1.13,~3.26,~2.36$ for $xy,~yz$ and $zx$ orbitals respectively. Here the orbital labelling is with reference to the octahedral directions characteristic of an RuO$_6$ octahedron. 

\section{Optical conductivity for $T=0$ single band Fermi liquid}

Here we give the details of the derivation of the $T=0$ optical conductivity obtained if the $T=0$ Fermi liquid self-energy is substituted into the Allen formula of the main text. We begin from the Allen formula, 
\begin{equation}
\sigma(\Omega)\propto\frac{i}{\Omega}\int d\omega\frac{f(\omega)-f(\omega+\Omega)}{\Omega-\Sigma(\Omega+\omega)+\Sigma^{\star}(\omega)},
\label{sigmaAllen}
\end{equation}
where the $\star$ denotes complex conjugation.

Now substituting the Fermi liquid self-energy [Eq.~\eqref{sigmafl}] into Eq.~\eqref{sigmaAllen} and taking the $T=0$ limit of the Fermi functions gives
\begin{equation}
\sigma(\Omega)\propto\frac{i}{\Omega}\int_{-\Omega}^0 d\omega\frac{1}{Z^{-1}\Omega-i\,\omo^{-1}\left(\omega^2+\left(\Omega+\omega\right)^2\right)}.
\end{equation}
Finally defining $\omega=\Omega x$  gives
\begin{equation*}
\sigma(\Omega)\propto\frac{i}{\Omega}\int_{-1}^0 \Omega dx\frac{1}{Z^{-1}\Omega-i\,\omo^{-1}\left(\Omega^2x^2+\Omega^2\left(1+x\right)^2\right)},
\end{equation*}
and rearranging gives the equation quoted in the text.

\section{Vertex Corrections}

Here we briefly discuss the vertex corrections in the $t_{2g}$ multiband situation considered here.

What is needed for vertex corrections to vanish is
\begin{enumerate}
\item {\em No on-site optical (dipole-allowed) transitions}: the optical absorption process requires moving an electron from one site to another.  This is trivially satisfied in the single-band model, but is also satisfied in multiband models of the type we consider where the interacting on-site orbitals are $d$-derived and all have  the same even parity.  If one were to consider a more general model that included say the Ru $4p$ or $5p$ orbitals, {\em and} were to include interactions linking the $3d$ and $4/5p$ orbitals, then vertex corrections would be non-vanishing. 

\item {\em Point parity symmetry}. The issue here is that if condition (1) is satisfied, the dipole required for the optical transition is made by moving an electron from one site to another. If moving the electron to the left or to  the right creates equal magnitude dipoles with opposite signs,  no on-site effect can be generated and vertex corrections do not arise.  
\end{enumerate}

Mathematically a vertex correction arises from $\delta \Sigma/\delta \vec{A}$; if the conditions above are satisfied then $\Sigma$ (which in single-site DMFT is site-local) is a scalar and there is no vector in the problem which can allow $\vec{A}$ to couple to it to first order.  Specifically, in the formalism used in our paper   the Green's function is a $12\times 12$ matrix. In the Wannier basis that we use, ($t_{2g}$-derived states localized on the 4 Ru atoms in the unit cell) the self-energy is independent of momentum in the Brillouin zone and is site-diagonal, so that the self-energy matrix is block (site) -diagonal in this basis. The bare conductivity operator has no matrix elements between states on the same site (within the same block): conductivity processes only transfer an electron from one site to another, via a matrix element which has odd parity. In these circumstances, the vertex correction for the optical conductivity indeed vanishes in the single-site approximation. 

In a crystal structure of sufficiently low symmetry, hybridization effects could in principle lower the point symmetry of a given Ru site, so that the on-site wave functions did not have pure $d$ character (referred to the Ru site), thereby creating a vector in the problem that would allow a non-zero $\delta \Sigma/\delta \vec{A}$; this would also activate an on-site optical transition, thereby requiring a vertex correction. However, we know that such effects are very small in the case we consider (to a very good approximation the four Ru in a unit cell are related by a translation and octahedral rotation and the octahedra are close to symmetric), and have been set to zero by the  approximation we are forced to make to set the off-diagonal terms in the hybridization function to zero. 

We further observe that additional small interband terms would only act to increase the differences between the actual conductivity and the Allen formula, strenghtening our conclusion that one cannot safely draw inferences about non-Fermi-liquid behavior or quasiparticle scattering rates from optical conductivities once the structure becomes non-cubic.

\end{bibunit}
\end{document}